\documentclass[iop,twocolappendix,floatfix]{emulateapj}

\usepackage{epsfig}
\usepackage{apjfonts}
\usepackage{amsmath,color,bm}
\usepackage{booktabs}
\usepackage[breaklinks,colorlinks,urlcolor=blue,citecolor=blue,linkcolor=blue]{hyperref}
 \usepackage[utf8]{inputenc} 
 \usepackage[T1]{fontenc}

\shorttitle{Constraints on compact dark matter from gravitational wave microlensing}
\shortauthors{Basak et al}

\usepackage{natbib}
\usepackage{nameref}
\usepackage{graphicx}
\usepackage{graphics}
\usepackage[space]{grffile}
\usepackage{latexsym}
\usepackage{amsfonts,amsmath,amssymb}
\usepackage{url}
\usepackage[utf8]{inputenc}
\usepackage{fancyref}
\usepackage{hyperref}
\usepackage{xcolor}
\usepackage[normalem]{ulem}

\def\mc{\mathcal}

\def\Hl{{\mc{H}_\ell}}
\def\Hu{{\mc{H}_\textsc{u}}}

\def\Blu{{\mc{B}_\textsc{u}^\ell}}

\def\zl{{z_\ell}}
\def\Pl{{P_\ell}}
\def\dTl{{\Delta T_\ell}}
\def\zs{{z_\textsc{s}}}

\def\Ml{{M_\ell}}
\def\Mlz{{M_\ell}^z}

\def\fdm{{f_\mathrm{DM}}}
\def\Omegadm{{\Omega_\mathrm{DM}}}

\def\umax{{u^\mathrm{max}}}
\def\LambdaL{{\Lambda_\ell}}
\def\LambdaLMax{{\Lambda_\ell}^\mathrm{max}}
\def\LambdaMax{{\Lambda}^\mathrm{max}}

\def\NL{{N_\ell}}
\def\ZL{{Z_\ell}}
\def\pL{{p_\ell}}

\def\flow{{f_\mathrm{low}}}
\def\Mcs{{\mathcal{M}_s}}

\def\MLz{{M_\ell^z}}
\def\hl{{h_\ell}}
\def\blambda{{\bm \lambda}}
\def\blambdal{{\bm \lambda}_\ell}

\begin{document}

\title{Constraints on compact dark matter from gravitational wave microlensing}

\author{S. Basak$^1$}
\author{A. Ganguly$^{1,2}$}
\author{K. Haris$^{1,3,4}$}
\author{S. Kapadia$^1$}
\author{A. K. Mehta$^{1,5}$}
\author{P. Ajith$^{1,6}$}

\affiliation{$^1$~International Centre for Theoretical Sciences, Tata Institute of Fundamental Research, Bangalore 560089, India}
\affiliation{$^2$~Inter-University Centre for Astronomy and Astrophysics, Pune 411007, India}
\affiliation{$^3$~Nikhef – National Institute for Subatomic Physics, Science Park, 1098 XG Amsterdam, The Netherlands}
\affiliation{$^4$~Institute for Gravitational and Subatomic Physics (GRASP), Department of Physics, Utrecht University, Princetonplein 1, 3584 CC Utrecht, The Netherlands}
\affiliation{$^5$~Max Planck Institute for Gravitational Physics (Albert Einstein Institute), D-14476 Potsdam-Golm, Germany}
\affiliation{$^6$~Canadian Institute for Advanced Research, CIFAR Azrieli Global Scholar, MaRS Centre, West Tower, 661 University Ave., Suite 505, Toronto, ON M5G 1M1, Canada}

\begin{abstract}
If a significant fraction of dark matter is in the form of compact objects, they will cause microlensing effects in the gravitational wave (GW) signals observable by LIGO and Virgo. From the non-observation of microlensing signatures in the binary black hole events from the first two observing runs and the first half of the third observing run, we constrain the fraction of compact dark matter in the mass range $10^2-10^5~{M_\odot}$ to be less than $\simeq 50-80\%$ (details depend on the assumed source population properties and the Bayesian priors). These modest constraints will be significantly improved in the next few years with the expected detection of thousands of binary black hole events, providing a new avenue to probe the nature of dark matter.  
\end{abstract}


\section{Introduction}
\label{sec:intro}


Astronomical observations have firmly established that a significant fraction of the mass energy in the Universe is in the form of \emph{dark matter}, which interacts only through gravity~\citep{Bertone:2016nfn}. Fundamental particles that are beyond the Standard Model of particle physics are the most popular candidate for dark matter. However, such particles have so far evaded a confident detection through direct or indirect methods~\citep{Roszkowski:2017nbc}. Massive astrophysical compact halo objects (MACHOs), in particular primordial black holes (PBHs), are also potential candidates of dark matter~\citep{Carr:2020xqk}. 

PBHs could be formed via the collapse of large overdensities in the early Universe. Their abundance is heavily constrained by the non-observation of their signatures in a variety of astronomical probes. These include the effects of Hawking evaporation, microlensing of stars and supernovae, gravitational wave (GW) observations, accretion effects from X-ray binaries, distortions of cosmic microwave background, dynamical effects such as the stability of certain wide binaries and stellar clusters, formation of large-scale structures, etc.~\citep{Carr:2020xqk, Carr:2020gox}. Nevertheless,  the possibility of them contributing to the dark matter cannot be ruled out in several mass windows. Recent observations of GWs from massive black hole binaries~\citep{LSC_2016firstdetection,LIGOScientific:2016dsl,LIGOScientific:2018mvr,LIGOScientific:2020ibl} have resulted in renewed  interest in PBHs~\citep{Sasaki:2018dmp,Carr:2020xqk}.

Here we present constraints on the abundance of MACHOs~\footnote{Although the prime candidates of MACHOs are PBHs, the microlensing effects are \emph{practically} the same for any compact object. Hence we keep the generic name MACHOs throughout this article.} through the non-observation of gravitational microlensing signatures in the GW signals detected by LIGO and Virgo. If a significant fraction of the dark matter is in the form of MACHOs in the mass range $\sim 10^2-10^5~{M_\odot}$, microlensing will introduce characteristic deformations on the GW signals produced by binary black hole mergers~\citep{2003ApJ...595.1039T,Jung:2017flg,Urrutia:2021qak}. Our search included the binary black hole events detected by LIGO and Virgo during their first (O1)~\citep{LIGOScientific:2016dsl} and second (O2)~\citep{LIGOScientific:2018mvr} observing runs as well as the first half of the third (O3a) observing run~\citep{LIGOScientific:2020ibl}. We use the non-observation of such signatures to constrain the fraction of dark matter in the form of MACHOs to be less than $\sim 50-80\%$. The precise constraints depend on the assumed source population properties and the Bayesian priors. While these  constraints are indeed modest, this method provides a  new way to probe the abundance of MACHOs in the high mass window. These constraints will  significantly improve in the next few years with the  detection of thousands of binary black hole events.

\section{Search for microlensing signatures in LIGO-Virgo binary black hole events}
\label{sec:search}

\begin{figure}[tbh]
 \centering 
\includegraphics[width=0.7\columnwidth]{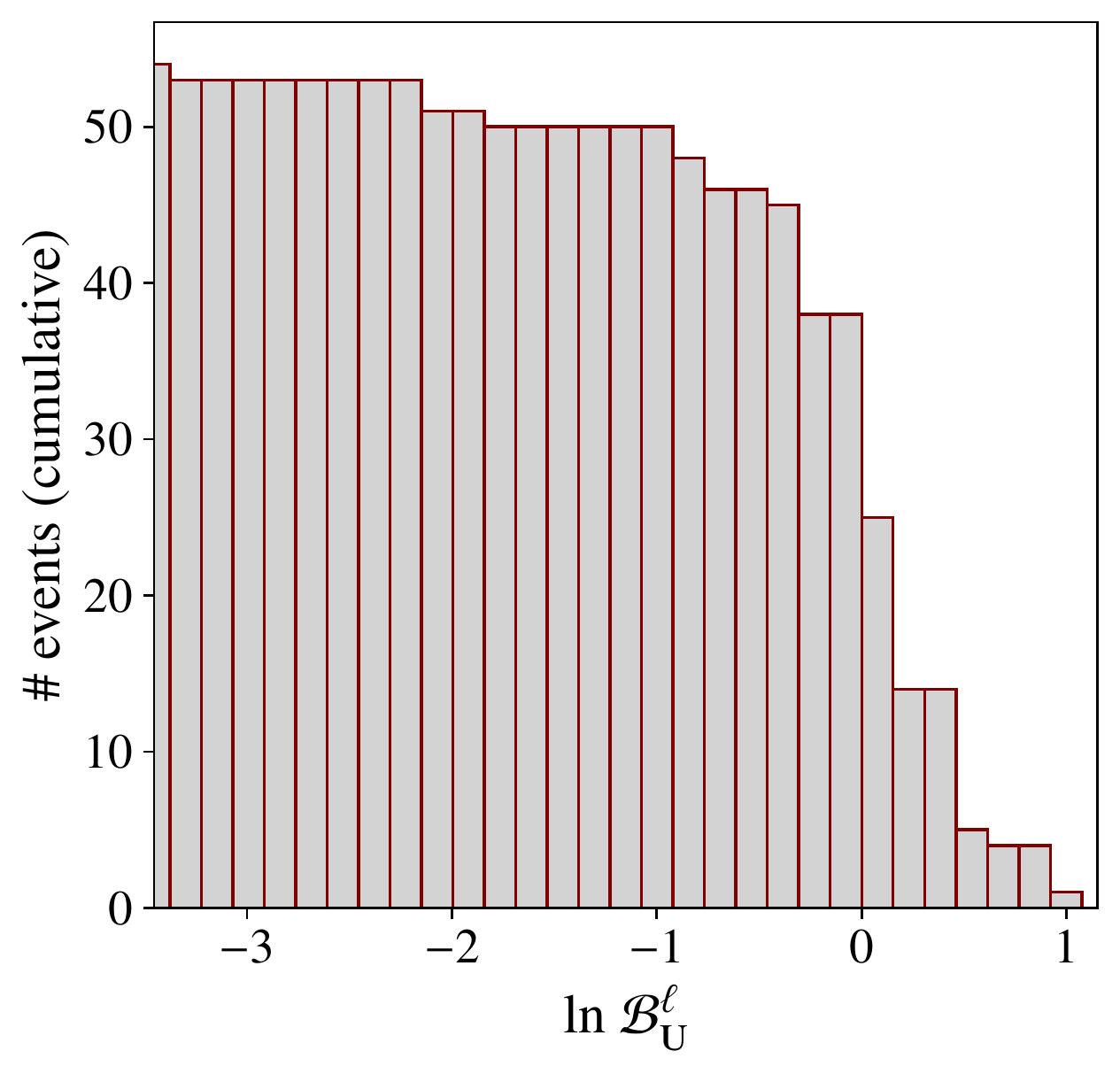}
\caption{Cumulative distribution of $\ln \Blu$ from LIGO-Virgo events from O1, O2 and O3a (number of events with $\Blu$ less than the value shown in the horizontal axis). The largest value is $\ln \Blu = 1.15$, which is not large enough to provide strong evidence for lensing.}
 \label{fig:BLU_dist}
\end{figure}

Microlensing of GWs involves qualitatively different features as compared to the microlensing of optical light from stars, supernovae, etc~\citep[e.g.]{Wyrzykowski:2009,EROS-2:2006ryy,MACHO:2004lxg}. Here, the wavelength of the radiation is comparable to the gravitational radius of the lens ($\lambda \sim G \MLz/c^2$). Hence wave diffraction effects will be apparent and the lensing has to be treated in the wave optics regime~\citep{2003ApJ...595.1039T}. 

Lensing effects on the GW signal $h(f; \blambda)$ (in Fourier domain) due to a point mass lens can be modeled in terms of a frequency dependent, complex magnification $F(f)$, so that the resulting lensed waveform is 
\begin{equation}
\hl(f; \blambda, \MLz, y) = h(f; \blambda) \, F(f; \MLz, y),
\label{eq:lens_waveform}
\end{equation}
where $\MLz \equiv \Ml (1+\zl)$ is the \emph{redshifted mass}\footnote{The frequency of GWs will be redshifted due to the cosmological expansion. This effect is equivalent to redefining the masses involved in the process $M \rightarrow M^z \equiv M(1+z)$ (both in the GW generation and lensing).} of the lens ($\Ml$ being its actual mass and $\zl$ the cosmological redshift) and $y$ is the dimensionless source position defined with respect to the optical axis~\citep{2003ApJ...595.1039T}. Also, $\blambda$ is the set of parameters that describe the (unlensed) GW signal in the detector, such as the redshifted masses ($m_1^z, m_2^z$), the dimensionless spin vectors ($\bm \chi_1, \bm \chi_2$), sky location of the binary ($\alpha, \delta$), luminosity distance ($d_L$), inclination and polarisation angles ($\iota, \psi$), and the time and phase of coalescence ($t_0, \phi_0$). 

Given the data $d$ containing a GW signal, and models of lensed and unlensed waveforms ($\Hl$ and $\Hu$), we can compute the Bayesian likelihood ratio between the ``lensed'' hypothesis $\Hl$ and ``unlensed'' hypothesis $\Hu$: 
\begin{equation}
\Blu = \frac{P( d | \Hl)} {P(d | \Hu)} = \frac{\int P({\blambdal} | \Hl) \, P(d | {\blambdal}, \Hl) \, d{\blambdal} } { \int P({\bm \lambda}| \Hu) \, P(d | {\bm \lambda}, \Hu) \, d{\bm \lambda} }, 
\label{eq:lens_bf}
\end{equation}
Where $\blambdal := \{\blambda, \MLz, y\}$ denotes the set of parameters describing the lensed waveform model. 

We searched for evidence of microlensing effects in the 10 binary black hole events reported by the LIGO-Virgo Collaboration from the first two observing runs~\citep{gwtc1}. Our search is similar to what is reported in~\cite{Hannuksela:2019kle}. However, we also include in our analysis, 8 additional events reported by~\cite{Zackay:2019tzo,Venumadhav:2019lyq, Zackay:2019btq}. The unlensed BBH waveforms $h(f)$ were generated using the \textsc{IMRPhenomPv2} waveform approximant~\citep{Hannam:2013oca,Husa_2016IMRPhenomD,Khan_2016IMRPhenomD} coded in the \textsc{LALSuite} software package~\citep{lalsuite}. We use the {Dynamic Nested Sampling~\citep{Dynesty:2020} implementation (\textsc{Dynesty}) in \textsc{Bilby} package \citep{Ashton:2018jfp} to compute the posteriors of the signal parameters and the marginal likelihoods of $\Hl$ and $\Hu$. In addition to this, we also make use of the results (that is, the $\Blu$ values) of the microlensing search on the 36 binary black hole events from the first half of the third observing run (O3a) reported by~\cite{LIGOScientific:2021izm}. 

For the Bayesian parameter estimation, we use uniform priors in the detector frame chirp mass $\mathcal{M}^z \in [3,60] ~{M_\odot}$ and the mass ratio $q \equiv m_2/m_1 \in [0.125,1]$, along with the constraint on the component masses $m_1^z, m_2^z \in [5,80] {M_\odot}$. We also use isotropic sky location (uniform in $\alpha, \sin \delta$) and orientation (uniform in $\cos \iota, \phi_0$), uniform in polarization angle $\psi$, and a volumetric prior $\propto d_L^2$ on luminosity distance. Additionally, we use a uniform prior in $\log_{10}~(\MLz/{M_\odot}) \in [0,5]$ and $p(y) \propto y$ with a cutoff $y \in [0.1, 3]$. In addition, we restrict the parameter space of lens parameters ($\MLz, y$) such that the time delay $\Delta t_\mathrm{lens}(\MLz, y)$ due to lensing is always less than the duration $\tau_\mathrm{signal}(\blambda)$ of the corresponding signal~\footnote{If the lensing time delay is larger than the duration of the waveforms, the resulting waveform will appear as two separate GW events in the LIGO-Virgo data. The non-observation of multiple images can also be used to put constraints on $\fdm$ at higher lens masses. This is being explored in an ongoing work.}. 

Figure~\ref{fig:BLU_dist} shows the distribution of $\Blu$ from the 54 binary black hole events detected during the O1, O2 and O3a. No event provides a strong support for the lensing hypothesis (largest $\ln \Blu$ being 1.15). We use this non-observation of lensing effects to put constraints on the fraction of compact objects forming dark matter. 

\section{Constraining the compact dark matter fraction from non observation of lensing effects}
\label{sec:constrain}

\begin{figure*}[tbh]
\centering
\includegraphics[width=2\columnwidth]{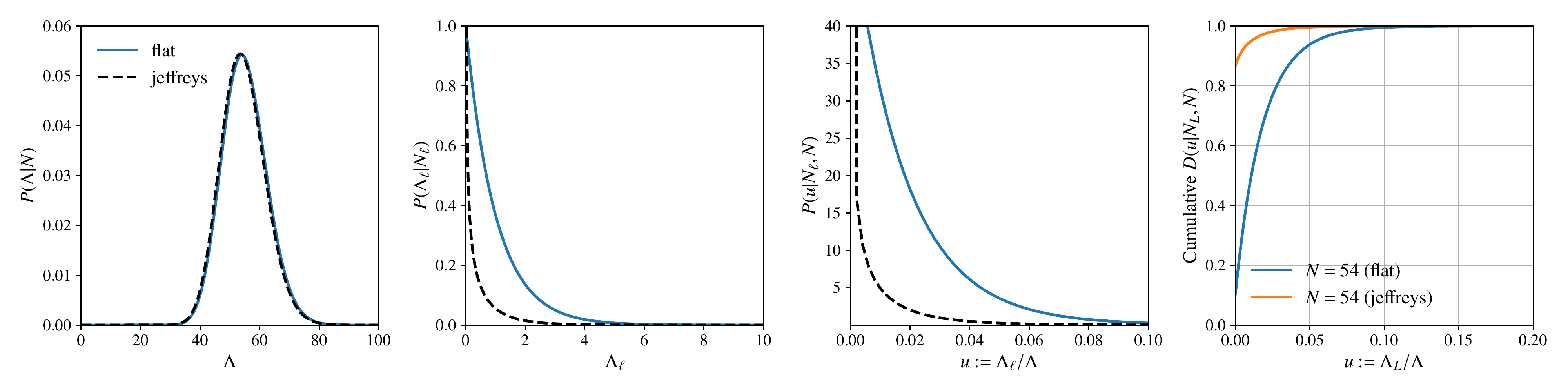}
\caption{Posterior distributions of the Poisson mean of the total number of detections ($\Lambda$), that of the number of lensed events ($\LambdaL$) and that of the fraction of lensed events ($u$) obtained from the O1, O2, O3a observation runs of LIGO and Virgo. Here the observed number of events $N = 54$ and the observed number of lensed events is $\NL = 0$.}
\label{fig:posteriors_Lambda_LambdaL_and_u}
\end{figure*}

Here we use the non-observation of lensing signatures to compute the posterior distribution of the fraction of lensed events among the detected events, and in turn, the posterior of the fraction of dark matter $\fdm$ in the form of MACHOs. 

We take that a total of $N = 54$ merger signals are confidently detected, and none of them are found to be lensed (i.e., $\NL = 0$). Further, we assume that the number of detected events follow a Poisson distribution with mean $\Lambda$, whose posterior distribution can be estimated as 
\begin{equation}
p(\Lambda | N) = Z^{-1} \, p(\Lambda) \, p(N | \Lambda),
\end{equation}
where $p(\Lambda)$ is the prior distribution on $\Lambda$ and $Z$ is the normalisation constant, while the likelihood is approximated by a Poisson distribution 
\begin{equation}
p(N | \Lambda) = \frac{\Lambda^N \, \exp(-\Lambda)}{N!}. 
\end{equation}
Similarly, from the observation of zero lensed events ($\NL = 0$), the posterior on the Poisson mean $\LambdaL$ of the number for lensed events can be calculated as 
\begin{equation}
\pL(\LambdaL | \NL = 0)  = \ZL^{-1} \, \pL (\LambdaL) \, \pL(\NL = 0 | \LambdaL), 
\end{equation}
where $\pL(\LambdaL)$ is the prior distribution on $\LambdaL$, and $\ZL$ is the normalisation constant. The likelihood is 
\begin{equation}
\pL(\NL = 0 | \LambdaL) = \frac{ \exp(-\LambdaL)} {1-\exp(-\LambdaLMax)},
\end{equation}
where $\LambdaLMax$ is the largest value that $\LambdaL$ can take (corresponding to the situation where all dark matter is in the form of MACHOs; i.e., $\fdm = 1$).

To compute the posterior on the fraction of lensed events $u \equiv \LambdaL / \Lambda$, we need to use the ratio distribution. This gives 
\begin{equation}
p(u \mid \{\NL = 0, N\}) \propto \int_0^{\infty} \frac{\Lambda^{N+1}}{\ZL} \pL(u \Lambda) p(\Lambda) \, e^{-\Lambda(u+1)}  d\Lambda,
\end{equation}
where the normalisation can be fixed by requiring  $\int_0^\umax p \left(u \mid \{\NL = 0, N\} \right) \, du  = 1$, where $\umax$ is the maximum possible value of $u$ (corresponding to $\fdm = 1$). 

Figure~\ref{fig:posteriors_Lambda_LambdaL_and_u} shows the posterior distributions of $\Lambda$, $\LambdaL$ and $u$ obtained from the the LIGO-Virgo events, assuming two different prior distributions for $\Lambda$ and $\LambdaL$. Finally, the posterior on $\fdm$ can be computed as 
\begin{equation}
p(\fdm \mid \{\NL = 0, N\})  =  p(u \mid \{\NL = 0, N\}) \left| \frac{d u}{d\fdm} \right|,
\label{eq:fdm_posterior}
\end{equation}
where $\frac{d u}{d\fdm}$ is the Jacobian of the lensing fraction $u$ and the compact dark matter fraction $\fdm$. 

We determine this Jacobian by simulating  astrophysical populations of binary black hole mergers and point mass lenses. We consider three different cosmological redshift distributions of binary black holes --- uniform distribution in comoving volume as well as that the ones predicted by  population synthesis models presented in~\cite{Dominik:2013tma} and \cite{Belczynski:2016obo,Belczynski:2016jno}. We use a power-law mass distribution model, $P(m_1)=m_1^{-2.35}$, on the mass of the heavier black hole while the mass ratio $m_2/m_1$ is distributed uniformly in the interval $[1,1/18]$ with the total mass lying in the interval $[5-200]~{M_\odot}$~\citep{LIGOScientific:2016kwr}. We consider spinning black holes with component spin magnitudes distributed uniformly between $0$ and $0.99$ with spins aligned/antialigned with the orbital angular momentum. The binaries are distributed uniformly in the sky with isotropic orientations.

In our simulations, MACHOs are approximated by point mass lenses and distributed uniformly in comoving volume, and the microlensing optical depth depends on the $\fdm$.  Lensing effects on the GW signal are computed using Eq.\eqref{eq:lens_waveform}. Binaries producing a network signal-to-noise (SNR) of 8 or above in the LIGO-Virgo detectors were deemed detectable. Since the calculation of the lensing likelihood radio $\Blu$ using Nested Sampling from all the simulated signals is computationally expensive, we use an approximation that is expected to be accurate in the high SNR regime~\citep{Cornish:2011ys,Vallisneri:2012qq}. We then compute the fraction of detected events that produce a $\Blu$ that is larger than the highest $\Blu$ obtained from real LIGO-Virgo events. This lensing fraction is shown as a function of the $\fdm$ in Fig~\ref{fig:lensing_frac_jacobian}. This data can be used to compute the Jacobian $\frac{d u}{d\fdm}$.

\begin{figure*}[tbh]
\centering
\includegraphics[width=\linewidth]{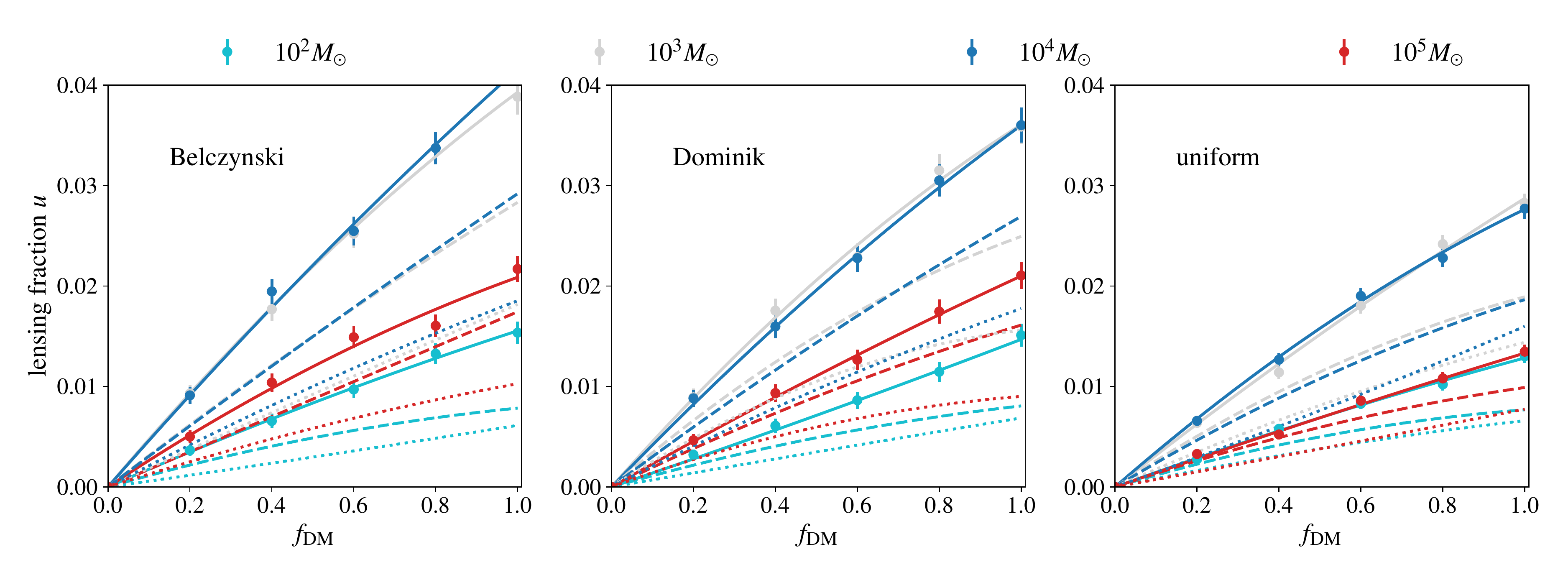}
\caption{The fraction of simulated events with $\ln \Blu$ greater than the threshold value 1.15 shown as a function of the $\fdm$. The left, middle and right plots correspond to different assumed redshift distribution of mergers. In each plot, different colors correspond to different lens masses (shown in legend). The solid, dashed and dotted lines correspond to the lensing fraction estimated using the noise power spectral densities of LIGO-Virgo detectors from O3a, O3 and O1 observing runs, respectively. The error bars indicate the counting errors due to the finite number of samples of simulated binaries and the curves show quadratic polynomial fits.} 
\label{fig:lensing_frac_jacobian}
\end{figure*}

\section{Results and discussion}
\label{sec:results}

\begin{figure*}[tbh]
\centering
\includegraphics[width=\linewidth]{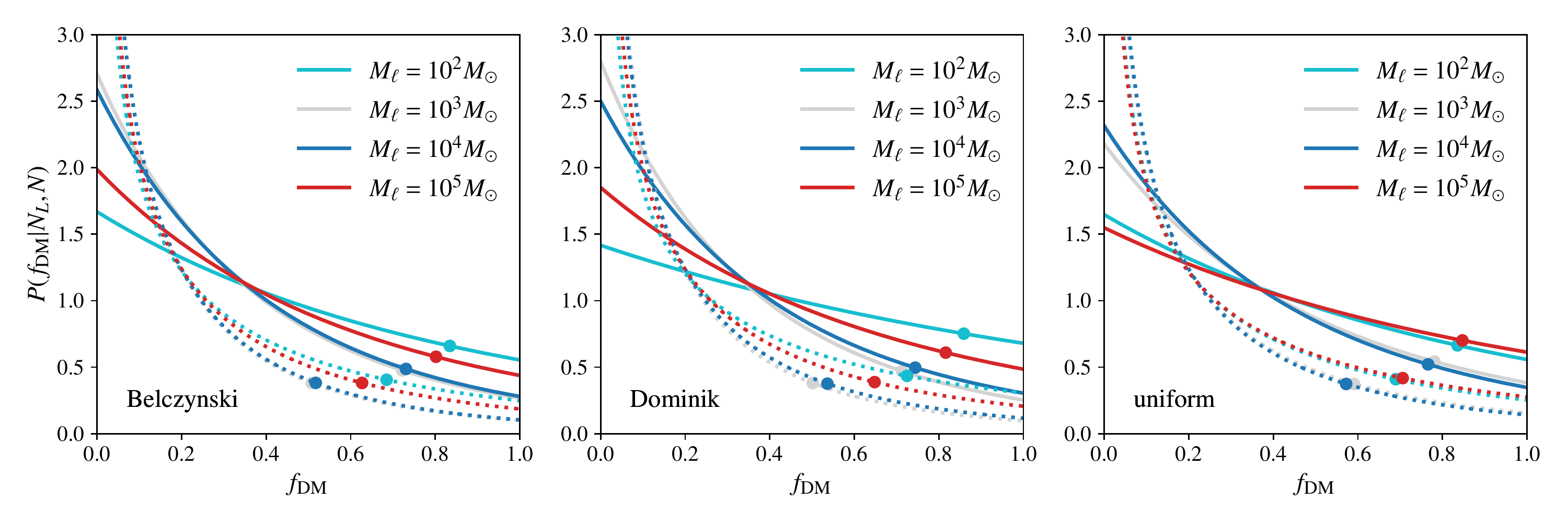}
\caption{Posteriors on $\fdm$ obtained by the non-observation of microlensing signatures in the 54 binary black hole events detected in O1, O2 and O3a. Posteriors shown by solid (dotted) lines are obtained by assuming flat (Jeffreys) prior in $\Lambda$ and $\LambdaL$. The left, middle and right plots correspond to different assumed redshift distribution models of binary black holes. In each sub-plot, different curves correspond to different assumed lens masses (shown in legends). The 90\% credible upper limits are shown by dots.}
\label{fig:fdm_posteriors}
\end{figure*}

\begin{figure*}[tbh]
\centering
\includegraphics[width=0.9\linewidth]{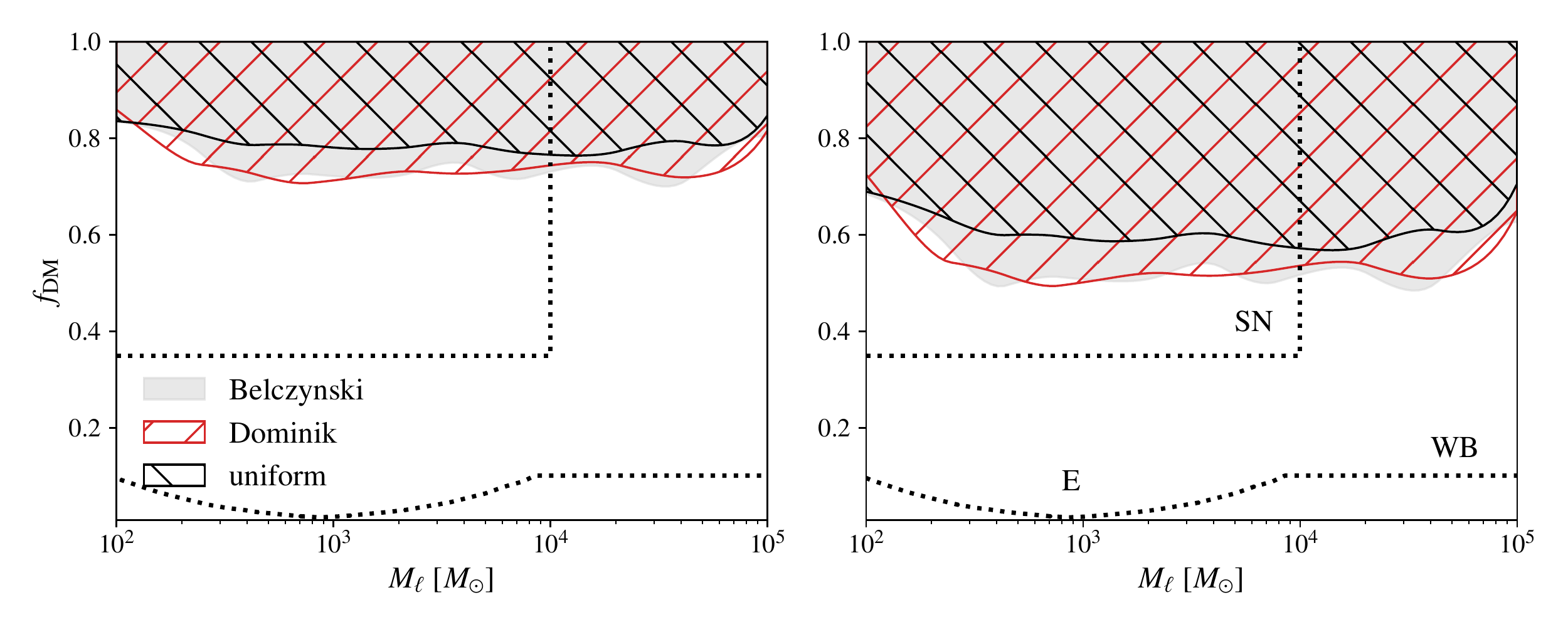}
\caption{90\% upper limits on $\fdm$ obtained from the O1, O2 and O3a events, assuming monochromatic mass spectrum for MACHOs (lens mass shown in the horizontal axis). The left (right) panel corresponds to bounds computed assuming flat (Jeffreys) prior on $\Lambda$ and $\LambdaL$. In each panel, three different exclusion regions correspond to three assumed models of the redshift distribution of binary black holes. The dashed lines show some of the the existing constraints from the microlensing of supernovae (SN) and from the stability of wide binaries (WB) and a star cluster in the galaxy Eridanus II (E)~\citep{Carr:2020xqk,Carr:2020gox}.} 
\label{fig:fdm_upperlimits}
\end{figure*}

\begin{figure}[tbh]
\centering
\includegraphics[width=0.95\columnwidth]{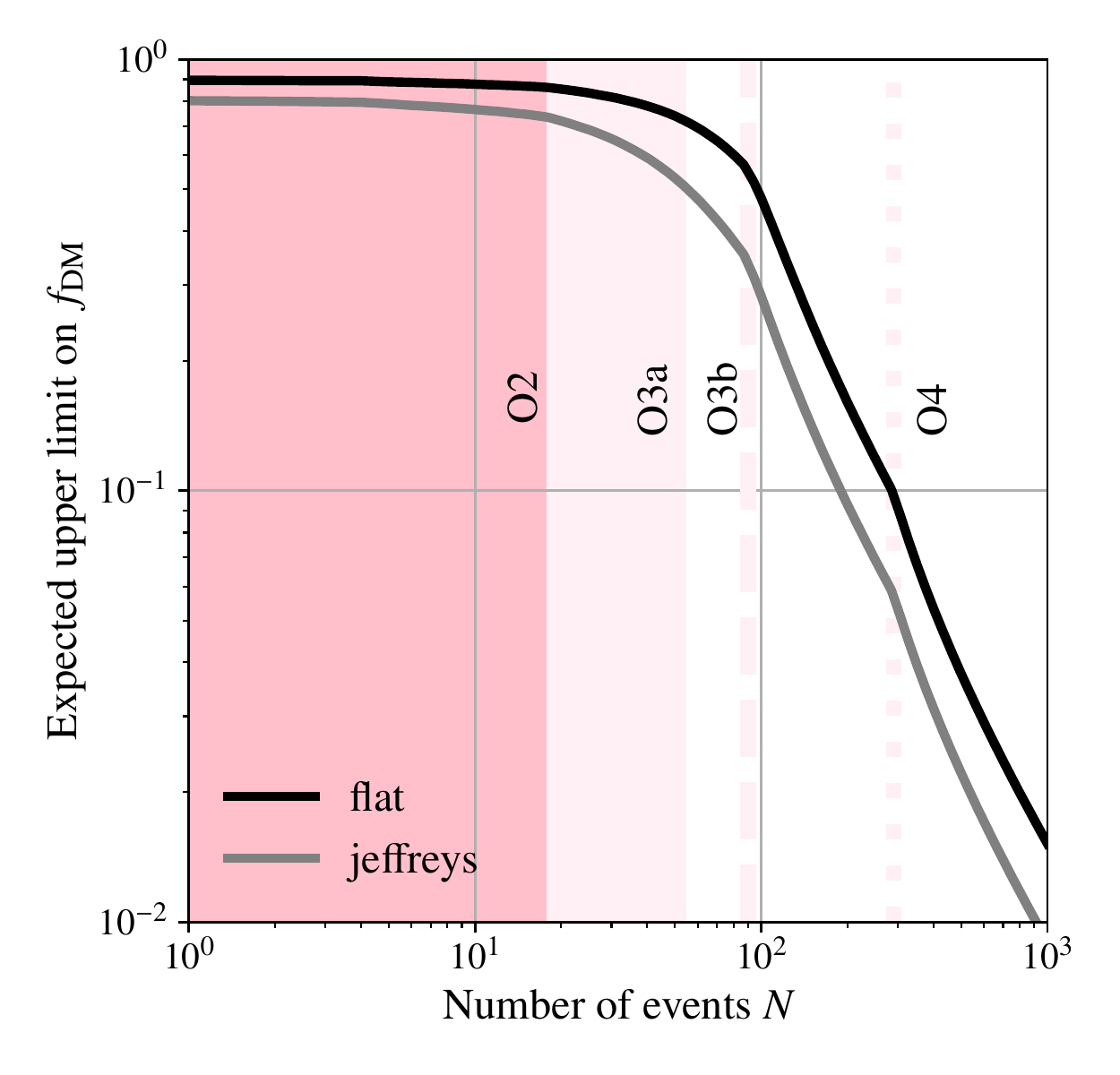}
\caption{Upper bounds on $\fdm$ expected from future observing runs, shown as a function of the cumulative number of detected binary black hole events (for lens mass $10^3 M_\odot$). The black (grey) curves show the bounds computed assuming flat  (Jeffreys) prior on $\Lambda$ and $\LambdaL$. The number of detected events in O2 and O3a are shown as vertical lines. We also show the approximate number of detectable events in the second half of the third observing run (O3b) and the fourth observing run (O4). The expected bounds fall faster with increased sensitivity anticipated in upcoming observing runs. We have used the redshift distribution of binary mergers given by \cite{Belczynski:2016obo} to compute these expected bounds.} 
\label{fig:fdm_bounds_projected}
\end{figure}


The largest value of the microlensing likelihood ratio obtained from O1, O2, O3a events is $\ln \Blu$ = {1.15}. The fraction of simulated events with $\ln \Blu \geq 1.15$ is shown as a function of the $\fdm$ in } Fig.~\ref{fig:lensing_frac_jacobian}, for different lens masses. This allows us to compute the Jacobian $du/d\fdm$ and thus the posterior on $\fdm$ as described by Eq.~\eqref{eq:fdm_posterior}. As commonly done in the literature, we assume monochromatic spectra for MACHOs~\citep{Carr:2020xqk}. Figure~\ref{fig:fdm_posteriors} shows the posterior of $\fdm$, with masses given in the legends\footnote{For lens mass  $\lesssim 100 M_\odot$,  lensing effects on the waveform are typically too weak to be identified (Fig.~\ref{fig:blu_scatter_ml_y_sim}).  If we estimate the $\fdm$ posteriors for these lens masses, we will be practically recovering the priors. Also, for lens mass $\gtrsim 10^5 M_\odot$, lensing time delays are typically large enough to produce multiple resolvable copies of the signals (geometric optics regime). This is why restricted the mass range to $10^2-10^5 M_\odot$. This window can be extended in the future, when the search sensitivities improve.}. The 90\% upper limits are shown as filled circles in each plot. The upper limits depend on the assumed redshift distribution of binary black holes as well as the Bayesian priors used in the analysis. Nevertheless, we are able to place upper bounds on $\fdm$ of the order of $50-80\%$. The 90\% upper limits are shown as a function of the lens mass in Fig.~\ref{fig:fdm_upperlimits}. 

We point out some limitations of our study: We assume that the GW signals are (possibly) lensed by only one microlens. However, if $\fdm \simeq 1$, a small number of sources at high redshifts ($z \geq 1.5$) could be potentially lensed by more than one lens. Even then, we expect the dominant lensing effect on the waveform will be due to one single lens. The loss of sensitivity of our search due to neglecting the  contributions of additional lenses is expected to be negligible. Also, in order to estimate the sensitivity of our search (or, the Jacobian of the lensing fraction and dark matter fraction), we use an approximation of the Bayesian likelihood ratio that is expected to be valid in high SNRs. While we expect this approximation to be reasonable for the SNRs that we consider, the quantitative effect of this needs to be verified by extensive  simulations. 


We approximate MACHOs as isolated point masses. Since these micro lenses are embedded in the lensing potential of the galaxy, the macro lens can cause additional effects when the micro lenses are within the Einstein radius of the macro lens~\citep[e.g.]{Cheung:2020okf}. This is especially important when the micro lenses are very close to the image locations of the macro lens, which is expected to happen only for a small fraction of MACHOs. We also neglect any additional effect of lensing by sub-structures in dark matter halos~\citep[e.g.]{Dai:2018mxx}. The clustering of MACHOs, which we neglect, is unlikely to change our results significantly~\citep{Zackrisson:2007bt}.

The bounds that we obtain are  weaker than some of the existing constraints~\citep{Carr:2020xqk,Carr:2020gox}. However, the GW lensing bounds will get significantly better in the next few years as the sensitivity of GW detectors improve. The sensitivity improvement will bring about two effects: Firstly, the increased number of total detections will allow us to estimate the lensing fraction $u$ better (see, e.g. Fig.~\ref{fig:posteriors_Lambda_LambdaL_and_u}). Secondly, the increased horizon distance of the detectors will increase the lensing optical depth and hence the fraction of lensed events (see, e.g., Fig.~\ref{fig:lensing_frac_jacobian}). The expected $\fdm$ upper limits from future detections~\citep{Aasi:2013wya} are shown in Fig.~\ref{fig:fdm_bounds_projected}  (for lens mass $10^3 M_\odot$) as a function of the number of detected binary black hole mergers, assuming that none of them show signatures of lensing. The upcoming third generation of GW detectors that will detect hundreds of thousands of binary black hole mergers every year by probing the high-redshift Universe ($z \simeq 15$), the constraints will improve by orders of magnitude. It is fair to say that microlensing of GWs is opening a powerful probe of the nature of dark matter. 

\section*{Acknowledgments} 
We are grateful to Anupreeta More for her careful review of the manuscript. We also thank Aditya Vijaykumar and the members of the LIGO-Virgo-KAGRA collaboration's lensing subgroup for useful discussions. Our research was supported by the Department of Atomic Energy, Government of India. SJK’s research was funded by the Simons Foundation through a Targeted Grant to the International Centre for Theoretical Sciences, Tata Institute of Fun- damental Research (ICTS-TIFR). PA's research was funded by the Max Planck Society through a Max Planck Partner Group at ICTS-TIFR and by the Canadian Institute for Advanced Research through the CIFAR Azrieli Global Scholars program. The numerical calculations reported in the paper were performed on the Alice computing cluster at ICTS-TIFR and the Sarathi cluster at IUCAA.
\newpage
\appendix

\section{Posteriors of lensing fraction assuming different priors}

Here we present the explicit expressions of the posteriors of the lensing fraction $u \equiv \LambdaL/\Lambda$ using different priors. Here $\Lambda$ and $\LambdaL$ are the Poisson means of the number of binary black hole detections and the lensing detections, respectively. If we assume flat priors for $\Lambda$ and $\LambdaL$; that is, 
\begin{eqnarray}
p(\Lambda) & = & \frac{1}{\LambdaMax} \, \Theta(\Lambda - \LambdaMax), \\
p(\LambdaL) & = & \frac{1}{\LambdaLMax} \, \Theta(\LambdaL - \LambdaLMax) = \frac{1}{ \Lambda \umax} \, \Theta(u - \umax), \nonumber 
\end{eqnarray}
where $\LambdaMax,  \LambdaLMax$ ans $\umax$ are the maximum possible values of $\Lambda, \LambdaL$ and $u$, respectively ($\umax$ corresponds to $\fdm = 1$). This results in the explicit expression 
\begin{equation}
p(u \mid \{\NL = 0, N\}) \propto  \Theta(\umax-u) \, \int_0^{\LambdaMax} \frac{\Lambda^{N+1} \, e^{-\Lambda(u+1)} }{1- e^{-\umax \Lambda}}d\Lambda. 
\end{equation}

On the other hand, if we assume Jeffreys prior for $\Lambda$ and $\LambdaL$; that is, 
\begin{eqnarray}
p(\Lambda) & = & \frac{1}{\sqrt{\Lambda \, \LambdaMax}} \,  \Theta(\Lambda - \LambdaMax), \\
p(\LambdaL) & = & \frac{1}{\sqrt{\LambdaL \, \LambdaLMax}}  \,  \Theta(\LambdaL - \LambdaLMax) = \frac{1}{ \Lambda} \frac{1}{\sqrt{u \umax}} \, \Theta(u - \umax), \nonumber 
\end{eqnarray}
this results in the explicit expression 
\begin{equation}
p(u \mid \{\NL = 0, N\}) \propto  \frac{\Theta(\umax-u)}{\sqrt{u}} \, \int_0^{\LambdaMax} \frac{\Lambda^{N} \, e^{-\Lambda(u+1)} }{\mathrm{erf}({\sqrt{\umax \Lambda}})}d\Lambda. 
\end{equation}
We use these expressions to compute the posteriors shown in Fig.~\ref{fig:posteriors_Lambda_LambdaL_and_u}. The normalisation is fixed by $\int_0^\umax p \left(u \mid \{\NL = 0, N\} \right) \, du  = 1$. 

\section{Astrophysical simulations of lensed mergers}
\label{sec:simulations}

\begin{figure}[tbh]
\centering
\includegraphics[width=0.85\columnwidth]{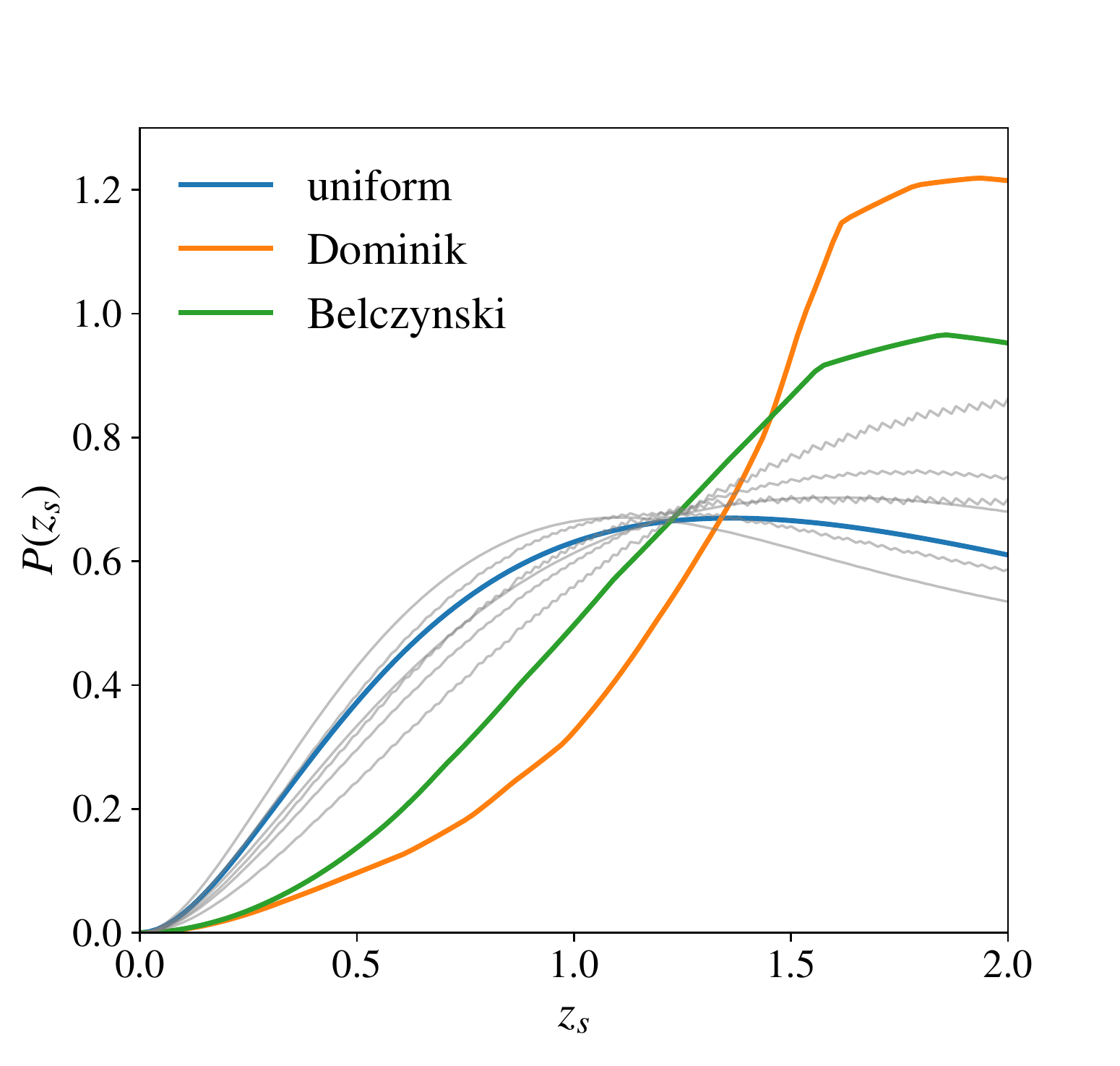}
\caption{Redshift distribution of binary black hole mergers assumed predicted by different models --- uniform distribution in comoving volume, population synthesis models predicted by \cite{Belczynski:2016obo,Belczynski:2016jno} and \cite{Dominik:2013tma}. We also show, in thin grey lines, several models of primordial black hole mergers given in~\cite{Mandic:2016lcn}. Since most of them are  ``bracketed'' by the three models that we consider, we do not use them explicitly in the computation of upper limits.}
\label{fig:merger_redshift_dist}
\end{figure}

\begin{figure*}[tbh]
\centering
\includegraphics[width=2\columnwidth]{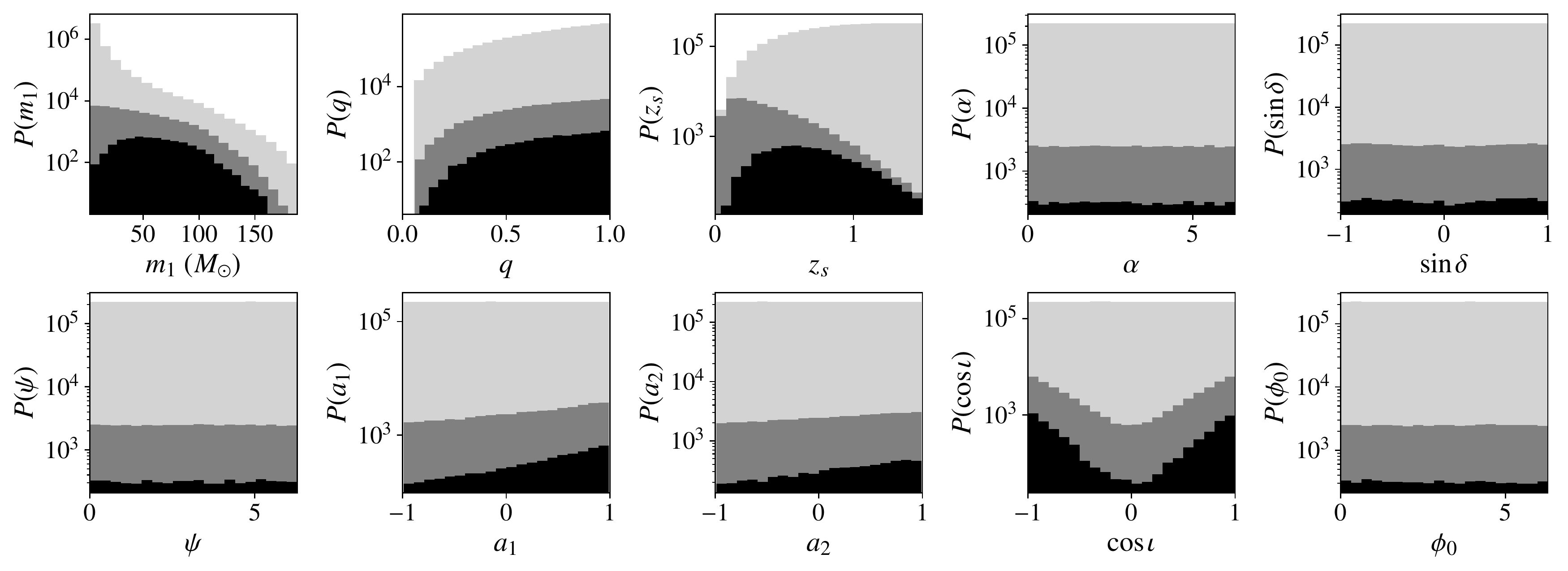}
\caption{Distribution of the simulated binaries (light grey), detected binaries (dark grey) and lensed binaries (black). The binaries are assumed to be distributed uniformly in comoving four-volume. The detector PSDs are from O3a.}
\label{fig:inj_dist}
\end{figure*}

\begin{figure}[tbh]
\centering
\includegraphics[width=0.85\columnwidth]{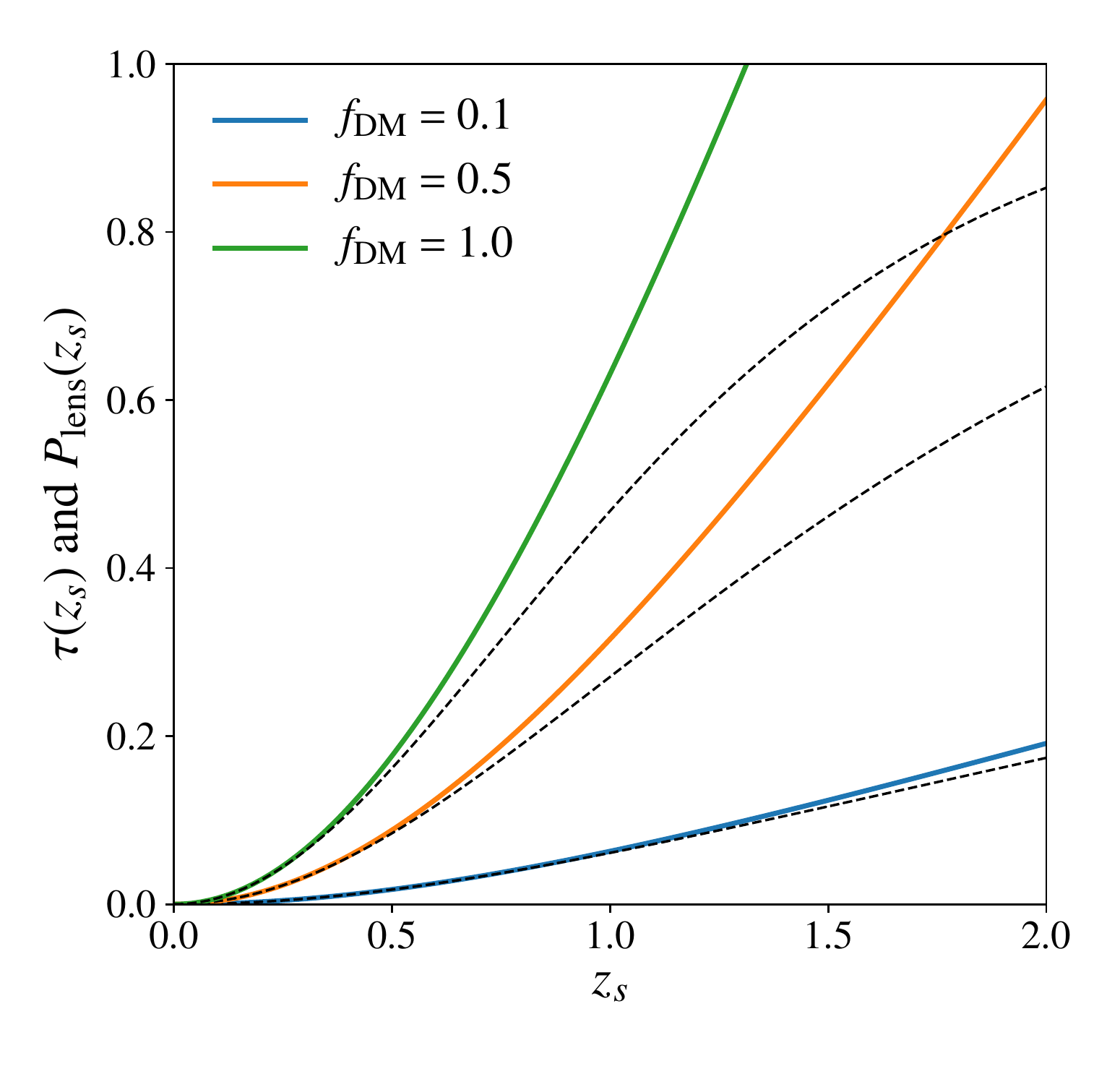}
\caption{Microlensing optical depth (solid lines) and lensing probability (dashed lines) as a function of the source redshift $\zs$ assuming different values of $\fdm$ (shown in legend).}
\label{fig:optical_depth}
\end{figure}

\begin{figure}[tbh]
\centering
\includegraphics[width=0.9\columnwidth]{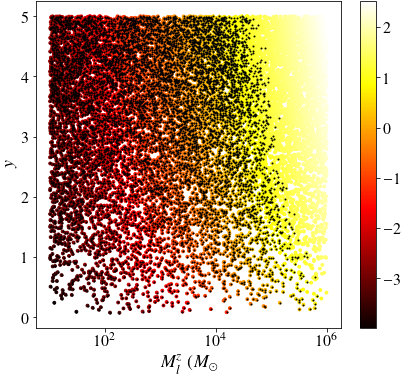}
	\caption{The color bar shows the lensing time delay $\log_{10} (\dTl/\mathrm{s})$ as a function of the redshifted lens mass $\Mlz$ and dimensionless impact parameter $y$. The black dots show the merger events for which $\dTl$ is less than the signal duration $\tau$.}
\label{fig:lensing_td_vs_M_y}
\end{figure}
\begin{figure}[tbh]
\centering
\includegraphics[width=\columnwidth]{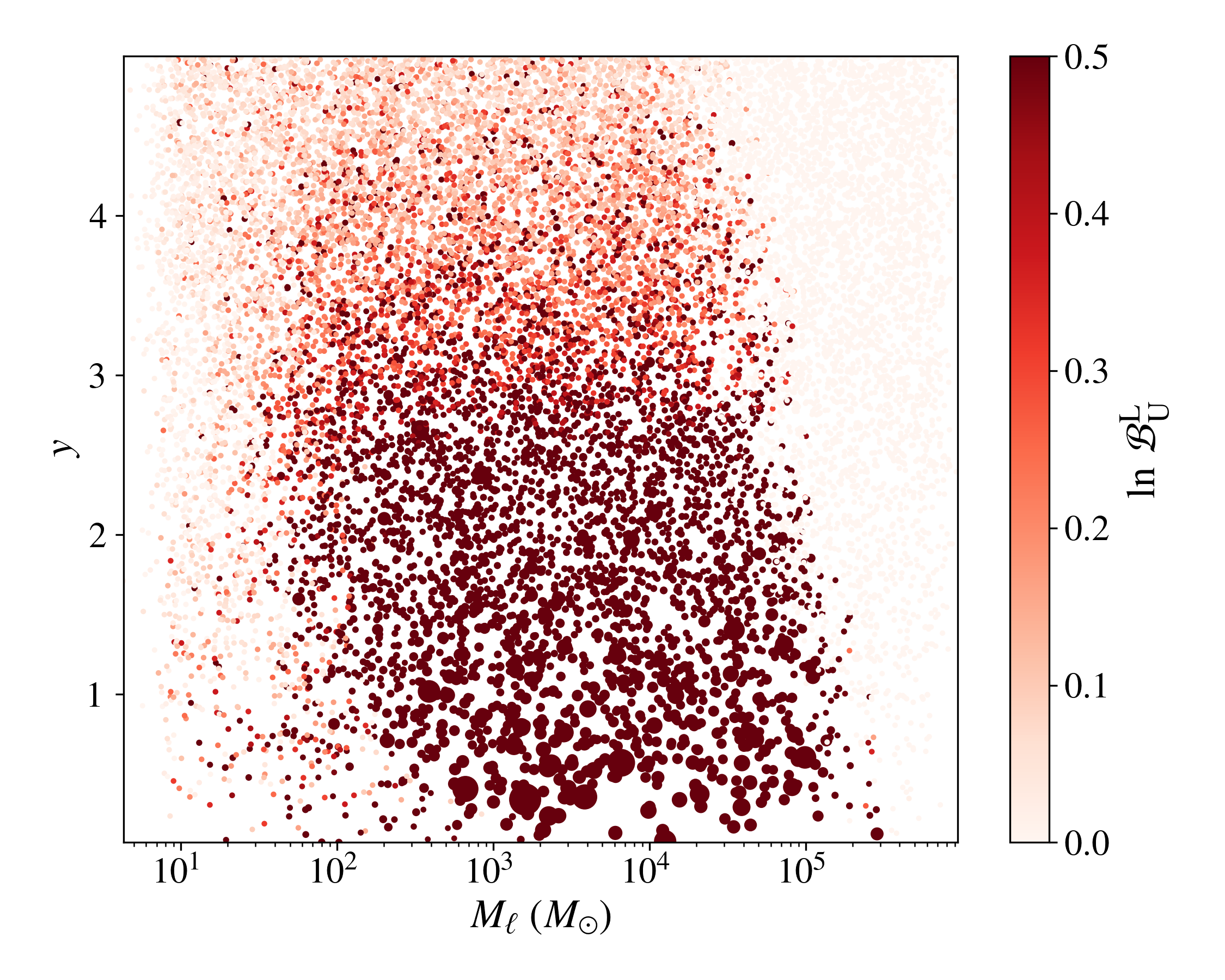}
\caption{Scatter plot of BBH events in the source frame lens mass $\Ml$ and impact parameter plane for $\fdm = 1$. The colour bar shows the value of the $\ln \Blu$. The simulation corresponds to a redshift distribution that is uniform in comoving four-volume.}
\label{fig:blu_scatter_ml_y_sim}
\end{figure}

Here we describe the astrophysical simulations used to evaluate the efficiency of our Bayesian model selection method in distinguishing lensed  merger events from unlensed events, thus estimating the Jacobian between the compact dark matter fraction $\fdm$ and the fraction of lensed events $u$. Here are the steps involved: 

\begin{enumerate}
\item \emph{Generate a population of mergers:} The source redshifts $z_s$ are drawn from three redshift distributions (Fig.~\ref{fig:merger_redshift_dist}) --- uniformly in comoving volume as well as from the population synthesis models given by~\cite{Dominik:2013tma} and \cite{Belczynski:2016obo,Belczynski:2016jno}. We use a power-law mass distribution model, $P(m_1)=m_1^{-2.35}$, on the mass of the heavier black hole while the mass ratio $m_2/m_1$ is distributed uniformly in the interval $[1, 1/18]$ with the total mass lying in the interval $[5-200]~M_\odot$. We consider spinning black holes with  component spin magnitudes distributed uniformly between $0$ and $0.99$ with spins aligned/antialigned with the orbital angular momentum. The binaries are distributed uniformly in the sky with isotropic orientations (Fig.~\ref{fig:inj_dist}).
\item \emph{Identify the detectable events:} Compute the optimal signal-to-noise ratio (SNR) of the binaries observed by the LIGO-Virgo detectors, using appropriate noise power spectral density (PSD) and antenna pattern functions. Binaries producing a network SNR of 8 or above are considered detectable (Fig.~\ref{fig:inj_dist}). 
A given choice of redshift and mass distribution of the mergers yields $\hat{N}$ number of detectable events. 

For O1, we used the representative PSDs given in~\cite{adligo-psd-o1-h1,adligo-psd-o1-l1}. For O2, the representative PSDs given in~\cite{H1L1V1-psd-O2} were used, while, for O3, O4 and O5 scenarios, the representative/anticipated PSDs given in~\cite{H1L1V1-psd-O3O4O5} were used. 
\item \emph{Identify the lensed events:} Assuming that the MACHOs are distributed uniformly in comoving volume, the probability that GWs from a binary located at a redshift of $\zs$ is lensed is given by $\Pl(\zs) = 1 - e^{-\tau(\zs)}$, where $\tau$ is the lensing optical depth (Fig.~\ref{fig:optical_depth})
\begin{equation}
\tau(z_s,\fdm,y_0) =   \int_0^{z_s} \frac{d\tau}{d\zl} ~ d\zl, 
\end{equation}
with the differential optical depth given by ~\cite{Jung:2017flg}
\begin{equation}
\frac{d\tau}{d\zl}  = \fdm ~ \frac{3}{2} y_0^2 \Omegadm \frac{H_0^2}{c} ~ \frac{(1+\zl)^2}{H(\zl)} \frac{D_{ls}D_l}{D_s}.
\label{eq:diff_opt_depth}
\end{equation}
Above, $y_0$ is a fiducial dimensionless radius of influence of the lens, or the maximum impact parameter (in units of the Einstein angle $\theta_E$) within which the lens can \emph{potentially} produce desired lensing effect~\footnote{As long as $y_0$ is chosen sufficiently large and the actual impact parameters are distributed up to $y_0$ (step 4), the precise choice of $y_0$ does not affect our estimation of the lensing fraction.}. 

We identify a binary as lensed when the lensing probability $\Pl(\zs)$ of that binary is larger than a random number uniformly distributed between 0 and 1. This ensures that $\Pl(\zs)$ fraction of binaries located at a redshift $\zs$ is counted as lensed. 
\item \emph{Assign lens properties:} When a merger located at a redshift $\zs$ is identified as lensed, the lens redshift $\zl$ is randomly drawn from a probability distribution given by the differential optical depth [Eq.~\eqref{eq:diff_opt_depth}]. For a lens mass $\Ml$, the redshifted lens mass is computed as $\Mlz = \Ml (1+\zl)$. The impact parameter $y$ is drawn from the distribution $P(y) \propto y$, with $y \in [0.01, y_0]$. We choose $y_0 = 5$, since signals with $y \gtrsim 5$ are  unlikely to contain identifiable lensing signatures (Fig~\ref{fig:blu_scatter_ml_y_sim})~\footnote{Since the optical depth is also scaled with the same value of $y_0$, this will  not change the fraction of identifiable lensed events.}. 
\item \emph{Identify events with wave optics effects:} Wave optics effects in the waveform are observed when the time delay caused by lensing is smaller than the duration of the signal. (Otherwise, lensing will produce multiple signals separated in time). The time delay produced by a point mass lens is given by~\citep{2003ApJ...595.1039T} 
\begin{equation}
\dTl = 4 \Mlz \left[ \frac{y\sqrt{y^2+4}}{2} + \rm {ln} \Big(\frac{\sqrt{y^2+4}+y}{\sqrt{y^2+4}-y}\Big) \right].  
\end{equation}
We approximate the duration of a GW signal by the Newtonian chirp time~\citep{Sathyaprakash:1994}, with some extra time to adjust for the presence of the merger and ringdown part. 
\begin{equation}
\tau_\mathrm{signal} = \frac{5}{256} {\Mcs^z}^{-5/3} (\pi \flow)^{-8/3} + 10^4 M_s^z, 
\end{equation}
where $\Mcs^z$ and $M_s^z$ are the redshifted chirp mass and total mass of the binary, respectively, while $\flow$ is the low-frequency cutoff of the detector. We consider those lensed binaries with $\dTl < \tau_\mathrm{signal}$ as the ones potentially containing wave optics effects (Fig.~\ref{fig:lensing_td_vs_M_y}).  
\item \emph{Generate lensed waveforms:} Generate gravitational waveforms corresponding to the source parameters. Apply the wave optics lensing effects using Eq.\eqref{eq:lens_waveform}. 
\item \emph{Compute the approximate Bayes factor for the microlensed events at each detector:} In the high SNR limit, the Bayes factor $\Blu$ between the lensed and unlensed hypotheses [Eq.~\eqref{eq:lens_bf}] can be approximated as~\citep{Cornish:2011ys,Vallisneri:2012qq},
\begin{equation}
 \ln \Blu \approx (1-\mathrm{FF}) ~ \rho^2,
\end{equation}
where $\rho \equiv \sqrt{(\hl | \hl)} \simeq  \sqrt{(h|h)}$ is the optimal SNR of the signal~\footnote{In the wave optics regime that we consider, the amplification of the signals is not substantial unlike in the geometric optics regime. Hence the approximation ${(\hl | \hl)} \simeq  {(h|h)}$ is a good one. The bias in  $\rho^2$ incurred by using this approximation is less than 10\% for over 90\% of the lensed signals.} while FF is the fitting factor of the unlensed waveform family $h(\Theta)$ with the lensed waveform $\hl$  
\begin{equation}
\mathrm{FF} = \max_{\Theta} ( \hl |h(\Theta))~.
\label{eq:ff}
\end{equation}
Here the brackets denote the following noise weighted inner product 
\begin{equation}
(a, b) = 4 \int_\flow^\infty \frac{a(f)b^*(f)}{S_h(f)} df 
\end{equation}
where $S_h(f)$ is the one sided power spectral density of the detector noise. In Eq.\eqref{eq:ff}, $\Theta$ comprises the intrinsic source parameters $\{\Mcs^z,\eta_s, \chi_s\}$ of the unlensed template. It is not necessary to maximize the match explicitly over the extrinsic parameters, as this is performed semi-analytically by the match calculation (for non-precessing signals containing only the dominant mode of the gravitational radiation) (Fig.~\ref{fig:blu_scatter_ml_y_sim}).
\item \emph{Combine Bayes factors from multiple detectors:} Assuming that the noise of different detectors are statistically independent, the Bayes factors $\Blu^{(D)}$ obtained from the individual detector $D$ can be combined as 
\begin{equation}
\Blu = \prod_D ~ \Blu^{(D)}. 
\end{equation}
\item \emph{Compute the fraction of detectable events that have a $\Blu$ greater than a threshold:} If an event has a $\Blu$ greater than a threshold, it is deemed as an event that is identifiable as lensed. If there are $\hat{N}_\ell$ such identifiable lensed events in the simulation, the {lensing fraction} $u$ is computed as $u \equiv \LambdaL/\Lambda \simeq \hat{N}_\ell/\hat{N}$, where $\hat{N}$ the  number of detectable events from the simulation. This lensing fraction $u$ as a function of $\fdm$ can be used to compute the Jacobian $du/d\fdm$ for a given choice of source population and lens mass (Fig.~\ref{fig:lensing_frac_jacobian}). 

Since the PSD of the detector noise is different between observing runs, this affects the fraction of lensed events for a given $\fdm$. We combine the lensing fraction $u(\fdm)$ computed from simulations using different PSDs, with the number of detected events from that observing run as the weight. That is, 
\begin{equation}
u(\fdm) =  \frac{1}{N} \sum_R N_{R} ~ u_R(\fdm), 
\end{equation}
where $N_{R}$ is the number of events detected (or, expected to be detected, in the case of future observing runs) in an observing run $R$ and $u_R(\fdm)$ is the lensing fraction estimated from simulations using the PSD of that observing run. $N$ is the total number of detected events considered. 

\end{enumerate}

\section{Bayesian model selection and injection studies}

\label{appn:pe_o1o2}

\begin{figure*}[tbh]
\centering 
\includegraphics[width=\textwidth]{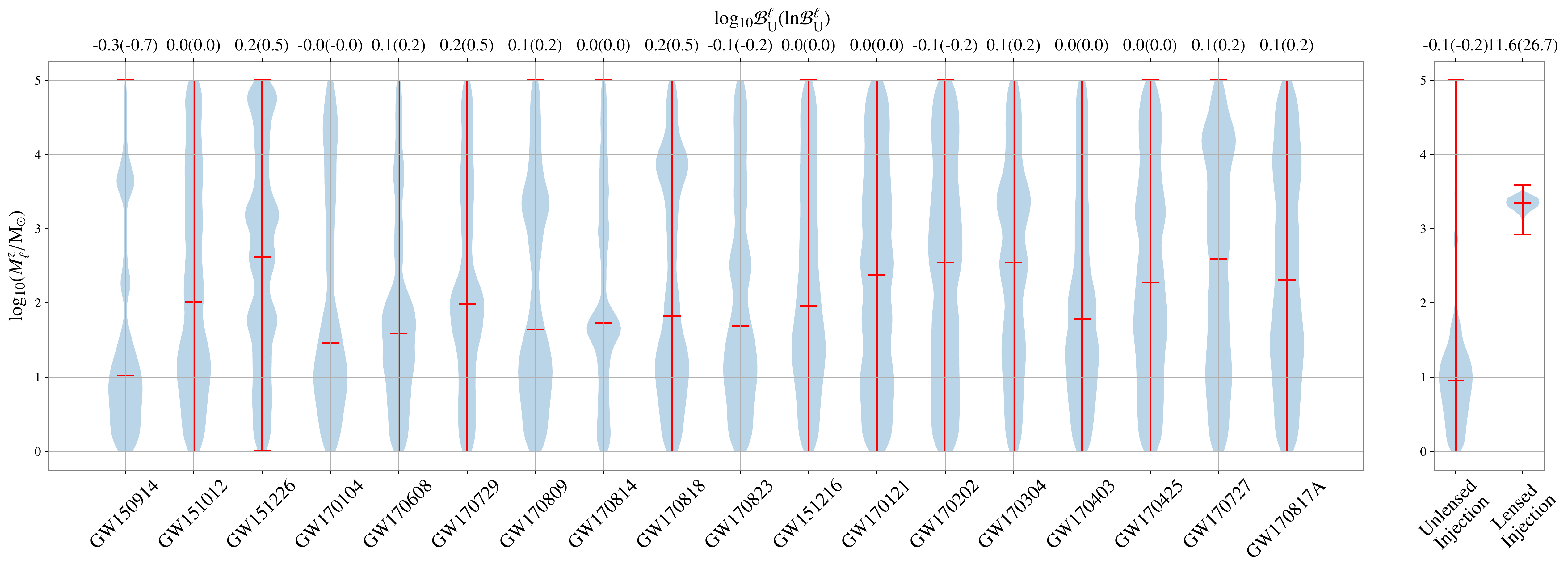}
\caption{The left panel shows the posterior distributions of redshifted lens mass $\Mlz$ (violin plots) and the Bayesian likelihood ratio between lensed and unlensed hypotheses (top horizontal axis) obtained from the binary black hole signals observed during O1 and O2. None of the likelihood ratios are significant enough to favor the lensing hypothesis. The right panel shows the same estimated from a simulated lensed/unlensed binary black hole event with redshifted masses ${m_1^z = 35.2 M_\odot, m_2^z = 31.7 M_\odot}$ (broadly consistent with the GW150914 event) and SNR = {16.1}. For the simulated lensed event, the lensing hypothesis is significantly preferred ($\ln \Blu = 26.7$) and the posterior on lens mass is consistent with the injected value $\Mlz = 10^{3.4} M_\odot$. For the simulated unlensed event,  $\ln \Blu = -0.2$  is consistent with the values derived from real events shown in the left panel (within noise induced fluctuations).}
 \label{fig:posteriors_O1O2}
\end{figure*}

Here we provide some additional details on the Bayesian model selection performed to compute the likelihood ratio $\Blu$, which, in turn, is used to determine whether a GW event contains signatures of microlensing or not. Figure~\ref{fig:posteriors_O1O2} (left panel) shows the posterior distributions of redshifted lens mass $\Mlz$ (marginalised over all other parameters) and the Bayesian likelihood ratio between lensed and unlensed hypotheses obtained from the binary black hole signals observed during O1 and O2. None of the likelihood ratios are significant enough to favour the lensing hypothesis. 

In order to check the accuracy of our Bayesian model selection, we perform a simulation study where an \emph{unlensed} GW signal with redshifted masses ${m_1^z = 35.2 M_\odot, m_2^z = 31.7 M_\odot}$ (broadly consistent with the GW150914 event) and SNR = {16.1} was added to Gaussian noise with the noise PSD from O3a. We then perform the Bayesian model selection using both the lensed and unlensed GW signal models. The true parameters of the simulated signal are well recovered within 90\% confidence interval of the posterior distribution of the parameters. The Bayesian likelihood ratio between lensed and unlensed hypotheses computed from this simulated event is $\ln \Blu = -0.2$, showing no evidence of lensing, as expected (Fig.~\ref{fig:posteriors_O1O2}  right panel). The recovered posterior on $\Mlz$ is  consistent with zero, as seen in the case of real events. We also simulate a signal with the same source parameters that is lensed by a compact object with redshifted mass $\Mlz = 10^{3.4} M_\odot$ and impact parameter $y = 0.47$ and repeat the same analysis on it. Here we find that the lensing hypothesis is significantly preferred ($\ln \Blu = 26.7$), as expected. The recovered posterior on $\Mlz$ is also consistent with simulated lens mass (Fig.~\ref{fig:posteriors_O1O2}  right panel). 

\bibliography{lensing_dm}

\end{document}